\documentclass[11pt,a4paper]{article}
\pdfoutput=1
\usepackage{graphicx,epsfig}
\usepackage{dcolumn} 
\usepackage{slashed,color,amsmath,amssymb}
\usepackage{jheppub}
\usepackage{enumitem}
\usepackage{mathrsfs}

\usepackage{caption, subcaption}
\usepackage{multirow}
\usepackage{verbatim}
\newcommand{\be}{\begin{equation}}
\newcommand{\ee}{\end{equation}}
\newcommand{\bea}{\begin{eqnarray}}
\newcommand{\eea}{\end{eqnarray}}
\newcommand{\bit}{\begin{itemize}}
\newcommand{\eit}{\end{itemize}}

\def\gsim{\lower0.5ex\hbox{$\:\buildrel >\over\sim\:$}}
\def\lsim{\lower0.5ex\hbox{$\:\buildrel <\over\sim\:$}}

\hypersetup{
   colorlinks=true,       
   linkcolor=blue,        
   citecolor=red,         
   filecolor=magenta,     
}

\usepackage{bm}

\preprint{}

\title{Mini-review on the production of Primordial Black Holes from First-Order Phase Transitions in the Early Universe}

\author{Indra Kumar Banerjee$^{1}$, Ujjal Kumar Dey$^{1}$, and Shaaban Khalil$^2$}

\affiliation{\vspace*{0.1in}$^1$ Department of Physical Sciences, Indian Institute of Science Education and Research Berhampur,Transit Campus, Government ITI, Berhampur 760010, Odisha, India}
\affiliation{\vspace*{0.1in}$^2$ Center for Fundamental Physics, \\
Zewail City of Science and Technology, 6th of October City, Giza
12578, Egypt}

\abstract{We review the creation mechanism of primordial black holes from first order phase transitions. We discuss various model-dependent and independent mechanisms and relate the properties of these mechanisms to the properties of primordial black holes. For each of these mechanisms, we provide model-specific examples.}

\begin{document}

\maketitle

\section{Introduction}

The early universe is believed to have undergone a series of dramatic phase transitions, driven by the cooling and expansion of the universe. These transitions played a crucial role in shaping the fundamental forces, particle content, and large-scale structure of the universe. Among them, First-Order Phase Transitions (FOPTs) are particularly significant, as they involve violent processes such as bubble nucleation, rapid expansion, collisions, and associated turbulence. These dynamics can lead to various interesting observables such as gravitational wave (GW) background, the baryon asymmetry, primordial black holes, Fermi balls, Q-balls, cosmic strings, etc. In this mini-review, we focus on the creation of primordial black holes from first order phase transitions.

Primordial black holes (PBH) are black holes formed in the very early stage of the universe~\cite{Zeldovich:1967lct}. These black holes are unique as they do not adhere to the usual low mass bounds that the astrophysical black holes maintain, i.e., the primordial black holes can have very low masses. This is specifically interesting in the context of Hawking radiation as the Hawking temperature, which is inversely proportional to the black hole mass, can be very high for these black holes. These PBHs could serve as a probe of high-energy physics and contribute to the dark matter density. The recent detection of GWs by LIGO, Virgo, and KAGRA~\cite{LIGOScientific:2025slb} has opened a new era of multi-messenger astronomy, primarily focusing on astrophysical sources such as black hole mergers and neutron star collisions. As a result, there is increasing interest in primordial black holes, including those originating from FOPTs in the early universe. Detection of these black holes could reveal physics beyond the Standard Model (SM), including:
\begin{itemize}
    \item Grand Unified Theories (GUTs),
    \item Electroweak symmetry breaking,
    \item Dark matter production mechanisms,
    \item Exotic phase transitions in hidden sectors.
\end{itemize}
Furthermore, if PBHs were abundantly produced during FOPTs, they could leave imprints in gravitational wave spectra, cosmic microwave background (CMB) anisotropies, and large-scale structure formation. Recent advances in numerical simulations and theoretical modeling have improved our understanding of bubble dynamics, particle production, and black hole formation during these transitions. All these make this an exciting frontier in cosmology.
This article provides a comprehensive review of the current understanding of PBH formation from FOPTs and their properties, depending on the specific creation mechanism. In Sec.~\ref{sec:pbhrecap} we provide a brief recap on the PBH formation mechanims and the constraints of PBH, followed by Sec.~\ref{sec:fopt} where we discuss the dynamics and the characteristic properties of FOPT. Next, in Sec.~\ref{sec:pbhfopt} we discuss the various model-independent and the model dependent mechanisms where PBHs can originate through a FOPT. In Sec.~\ref{sec:example} we provide examples of various work which use these PBH creation mechanisms using extensions of standard model and finally in Sec.~\ref{sec:conc} we conclude.

\section{Primordial Black Hole: A brief recap}
\label{sec:pbhrecap}

\begin{figure}
    \centering
    \includegraphics[scale=0.7]{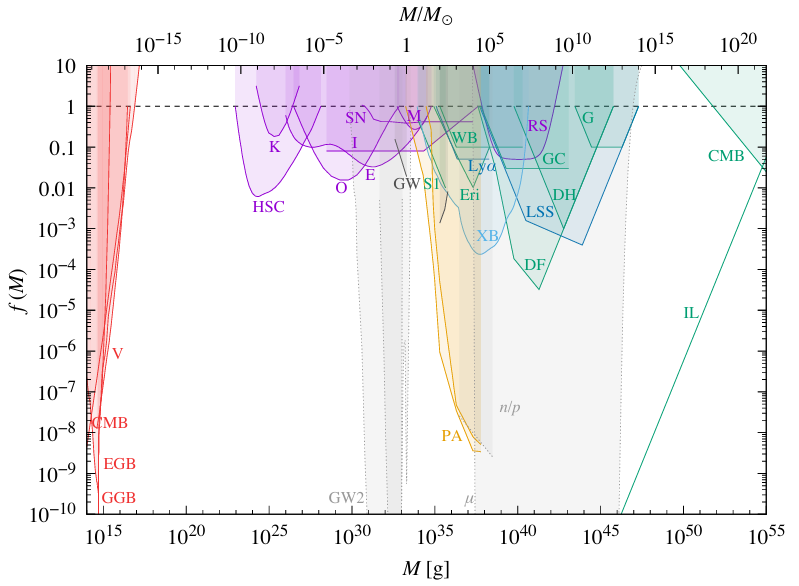}
    \caption{The constraints on PBH abundance $f(M)$ depending on their mass $M$. This figure is taken from Ref.~\cite{Carr:2020gox}.}
    \label{fig:PBHcarr}
\end{figure}

Primordial black holes are hypothesized as black holes formed in the early universe. Similar to astrophysical black holes, PBHs can also only be indirectly observed. However, unlike astrophysical black holes PBHs can be of extremely tiny mass. However, if the mass of a PBH is very tiny, then they likely have vanished through Hawking evaporation. Hence, the lightest non-evaporated PBH can have a mass $\sim 5\times 10^{14}\mathrm{~g}$~\cite{Carr:2020gox}.
Now we discuss the scenario of PBHs as the dark matter in the universe. It is speculated with evidence that the dark matter in the universe has to be non-baryonic, preferably cold, and interacts either very weakly or only gravitationally with visible matter. PBHs, which are speculated to have originated in the very early stage of the universe, satisfy all these conditions and hence can be viable dark matter candidates. However, observational data indicate that the relative density of the cold dark matter in the universe, $\Omega_{\mathrm{DM}} = 0.26$~\cite{Planck:2015fie, Planck:2018vyg}, that is approximately $26\%$ of the energy budget of the universe is dark matter. Therefore, one usually uses the abundance of PBHs, $f_{\mathrm{PBH}}$, which denotes the fraction of dark matter that is in the form of PBH. This, in a way, also quantify the density of the PBH in the universe today.
Based of the lack of direct and indirect observational signals, bounds have been put on PBHs in the form of the highest allowed abundance in the PBH mass-abundance plane and have been shown in Fig.~\ref{fig:PBHcarr}. 
The different constraints shown in the figure are explained briefly.
\begin{enumerate}
    \item \textbf{Overproduction of PBH:} The horizontal dashed black line denotes $f_{\mathrm{PBH}} = 1$, which suggests that all the dark matter in the universe is in the form of PBHs. In the range $10^{17}\mathrm{~g}\lesssim M_{\mathrm{PBH}} \lesssim 10^{22}\mathrm{~g}$, this is the only constraint on PBH abundance.
    \item \textbf{Evaporation constraint:} Hawking evaporation is a phenomenon through which black holes radiate particles of all species democratically depending on the mass of the black hole and the particles. Hawking evaporation can also be used to put bounds on PBH abundance. The Hawking temperature of a black hole ($T_{\mathrm{H}} = 1/8\pi G M_{\mathrm{BH}}$ where $\hbar = c = k_B =1$) is inversely proportional to the mass of the black hole, and therefore, these bounds are only applicable to PBHs of very light masses, shown with red curves. These bounds have been estimated through the observational constraints from CMB anisotropies~\cite{Carr:2009jm,Zhang:2007zzh,Adams:1998nr}, $\gamma$-ray background from galactic~\cite{EGRET:1997qcq,Wright:1995bi} and extra-galactic~\cite{MacGibbon:1991vc,Carr:1998fw} sources, and the Voyager-I electron-positron data~\cite{Boudaud:2018hqb}.
    \item \textbf{Microlensing constraints:} 
    Since the PBHs are compact objects, theoretical predictions (and non-observation) of microlensing events have also been used to put bounds on PBH abundance, which have been shown with magenta curves. These bounds come from the microlensing of various sources such as stars in the local neighborhood~\cite{MACHO:2000bzs, MACHO:2000qbb}, M31 galaxy~\cite{POINT-AGAPE:2005swi}, Magellanic clouds~\cite{Paczynski:1985jf, Wyrzykowski:2010mh, Wyrzykowski:2011tr}, etc., along with lensing of supernova~\cite{Metcalf:2006ms, Garcia-Bellido:2017imq}, quasars~\cite{Mediavilla:2009um, Mediavilla:2017bok}, etc. 
    \item \textbf{Dynamical constraints:} If PBHs form a significant part of the universe, then they would also take part in various events of dynamical nature, such as collision or encounter with earth leading to seismic activities~\cite{Khriplovich:2008er,Luo:2012pp}, encounter and merger with stars and other compact objects leading to visible $\gamma$-ray bursts~\cite{Zhilyaev:2007rx,Roncadelli:2009qj,Abramowicz:2008df}, encounter with wide binary star systems leading to the disruption of the binary system~\cite{Bahcall:1984ai,1987ApJ...312..367W,Yoo:2003fr,Quinn:2009zg} etc. The observational bounds from these predictions are shown with green curves. 
    \item \textbf{Constraints from structure formation:} Constraints on PBH from BBN state that the PBHs had to be produced before the universe was 1s old. This leads to the issue that these PBHs, if large enough and abundant enough can lead to early structure formation~\cite{Carr:2018rid}. Observational bounds from the formation of various types of galaxies and galaxy cluster therefore put bounds on the abundance of heavy PBHs~\cite{10.1093/mnras/206.4.801,Meszaros:1975ef,1977A&A....56..377C}, which are shown with dark blue curves. 
\end{enumerate}
For the detailed explanation of these mechanisms and all the other constraints, the reader is encouraged to go through Ref.~\cite{Carr:2020gox} and references therein. 


The main idea behind the creation of PBH is density (or curvature) perturbations and their collapse. A very brief and simplified description of the PBH creation is as follows. In general, it is considered that in the very early stage, the universe was highly homogeneous and isotropic, barring some perturbations. Therefore, one assumes an average density of the homogeneous and isotropic background as $\rho_b$. If a particular region has a density $\rho(x)$, then one can calculate the density contrast as, $\delta(x) = (\rho(x)-\rho_b)/\rho_b$. If the density contrast of a certain region exceeds a threshold value, then one can conclude that the region collapses to form a PBH. Now, various studies have shown that this threshold $\delta_c \in [0.4,0.67]$~\cite{Musco:2018rwt}. There are various mechanisms that may lead to such a density contrast. The most studied PBH formation mechanism is the collapse of overdense regions originating from the curvature perturbations created during inflation. During inflation, curvature perturbations of various length scales originate from the quantum fluctuations of the existing fields, which are stretched during the universe's exponential expansion. The small-scale curvature perturbations oscillate and eventually die down, whereas the perturbations of length scale larger than the Hubble radius stay frozen in space. However, as the universe expands post-inflation, these large-scale perturbation re-enters the horizon. If the amplitude of these perturbations exceeds a certain threshold, they can create overdense regions that may eventually collapse to form primordial black holes. 
In this mini-review, we discuss scenarios of PBH creation from first order phase transitions. Recently there have been studies which shows that there exist various creation mechanisms of PBH from FOPT. Some of these mechanisms are model-independent, whereas some only works for particular models. However, all these mechanisms revolve around the collapse of regions created directly or indirectly due to a FOPT. Before discussing these mechanisms, we first briefly describe dynamics of cosmological first order phase transitions.

\section{First-Order Phase Transition}
\label{sec:fopt}
A first-order phase transition (FOPT) is a type of phase transition in which a system undergoes a sudden and discontinuous change in its physical properties, such as density, magnetization, or molecular order, as a result of a change in external conditions like temperature, pressure, or magnetic field.  FOPT in cosmology is particularly significant, as it is characterized by the presence of a metastable state and a discontinuous change in the order parameter~\cite{Coleman:1977py, Callan:1977pt}. 
\begin{figure}[h]
    \centering
    \includegraphics[width=0.5\linewidth]{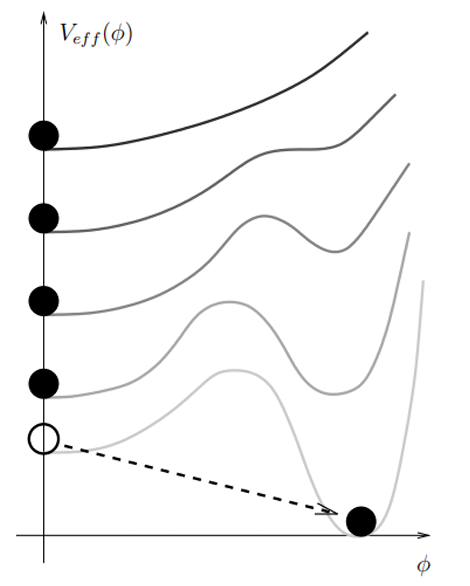}
    \caption{Evolution of the effective potential with temperature, where the upper curves denote the potential at higher temperature. The figure is taken from Ref.~\cite{Rubakov:2014erg}.}
\label{fig:fopt_pot_scheme}
\end{figure}
The dynamics of the universe can be expressed by the effective temperature-dependent potential, which is shown by a schematic diagram in Fig.~\ref{fig:fopt_pot_scheme}. At very high temperature, there is symmetry as the minima is at $\phi = 0$, where the $\phi$ is the classical background of some relevant scalar field that drives the FOPT. As the temperature reduces, the potential develops a kink, and eventually, after a certain temperature, the potential realizes a new global minima which is separated from the previous local minima by a barrier. The presence of this barrier determines whether a phase transition is first-order or second-order. Now the universe shifts from the false vacuum (local minima at $\phi = 0$) to the true vacuum through the means of thermal tunneling in the temperature-dependent case. Physically, this corresponds to the nucleation of bubbles of true vacuum. Once the true vacuum bubbles are stable, they release latent heat, which is due to the energy density difference between the false and true vacuum. Some part of this energy is transferred to the walls of the true vacuum bubbles, leading to the expansion of the bubbles until they collide with other true vacuum bubbles, eventually shifting the entire universe to the true vacuum state. This entire process is shown through a few frames of a simulation~\cite{Weir:2017wfa}.

\begin{figure}
    \centering
    \includegraphics[width = 0.8\linewidth]{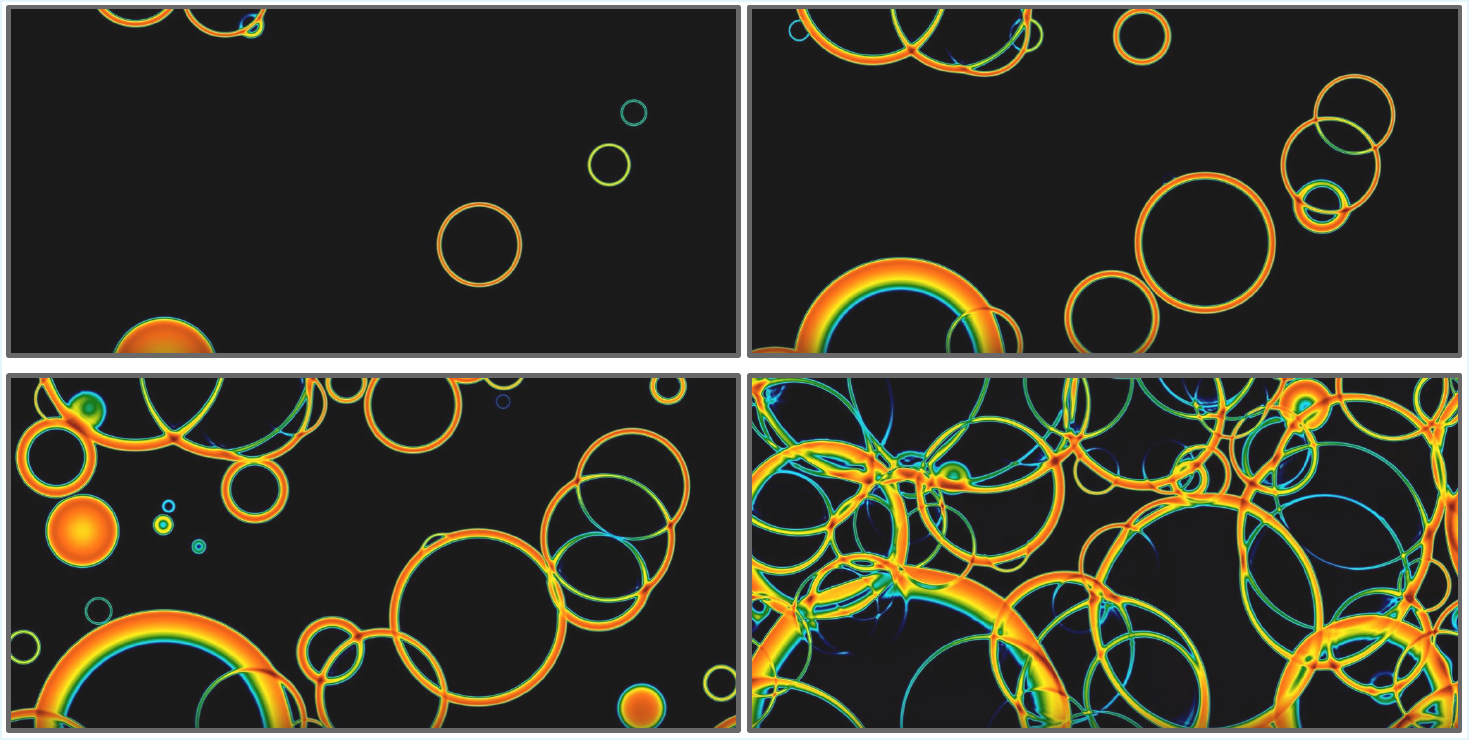}
    \caption{The evolution of a patch of the universe during a FOPT. The evolution of the patch proceeds as Top-left $\rightarrow$ Top-right $\rightarrow$ Bottom-left $\rightarrow$ Bottom-right. The figure is constructed from frames of a simulation performed in Ref.~\cite{Weir:2017wfa}. }
    \label{fig:bub_scheme}
\end{figure}
It can be seen from Fig.~\ref{fig:bub_scheme} that initially the very few bubbles are nucleated and they are all disconnected, and as the universe cools down, the bubbles expand and they collide eventually tracing more and more volume, facilitating the completion of the phase transition.


In order to study the details of the scalar-driven FOPT, we first express the full one-loop corrected potential $V_{\text{tot}}$ as follows,
\begin{align}
V_{\text{tot}} = V_{\text{tree}} + V_{\text{CW}} + V_{\text{th}} ,
\end{align}
where $V_{\text{tree}}$ represents the tree-level potential of a scalar field, whereas $ V_{\text{CW}} $ is the one-loop Coleman-Weinberg potential, and $V_{\text{th}}$ is the thermal correction to the potential. It is to be noted that although the tree-level potential signifies the basic shape of the potential and usually determines the zero-temperature minima of the system, it can not capture two very important effects, i.e., (i) the quantum corrections, and (ii) the finite temperature correction. 
Since quantum field theory is the conceptual backbone of microphysics that leads to the FOPT, one needs to consider the loop corrections as the scalar that drives the FOPT generally couples to other fields and hence the quantum correction can have important theoretical implications, such as shifting the minima, inducing spontaneous symmetry breaking, etc. The Coleman-Weinberg potential incorporates these quantum effects at the one-loop level~\cite{Coleman:1973jx}.
On the other hand, when considering the early universe, one must take finite temperature effects into account, which are encoded in $V_{\text{th}}$. These finite-temperature effects can lead to thermal corrections to the masses of the fields, potentially creating a barrier between the true and false vacuum, etc. The presence of the thermal correction also introduces temperature-dependent effects, which in turn leads to a potential that has temperature-dependent evolution. 
The thermal potential $V_{\text{th}}$ is given by,
\begin{align} 
V_{\text{th}}(\phi,T) = \frac{T^4}{2\pi^2}
	\left[ 
	\sum_{i=\text{bosons}} 
	J_{B}\left(\frac{m_i(\phi)}{T}\right) 
	+ \sum_{i=\text{fermions}} 
	J_{F}\left(\frac{m_i(\phi)}{T}\right)
	\right] 
	+ V_{\mathrm{D}}(\phi,T).
\end{align}
where the sums run over all bosonic and fermionic degrees of freedom, and $J_B $ and $ J_F $ are thermal loop functions that depend on the mass of the particles and the temperature $ T $. These functions can be expressed as,
\begin{align}
    J_B(x) &= \int_0^{\infty} dy~y^2 \ln\left[1-e^{-\sqrt{x^2+y^2}}\right],\\
    J_F(x) &= \int_0^{\infty} dy~y^2 \ln\left[1+e^{-\sqrt{x^2+y^2}}\right].
\end{align}
Here, $V_{\mathrm{D}}$ is the contribution from the daisy resummation and can be expressed as~\cite{Arnold:1992rz},
\begin{align}
V_{\mathrm{D}}(\phi,T)=-\sum_{i}\dfrac{g_iT}{12\pi}\left[m_i^3(\phi,T)-m_i^3(\phi)\right],
\end{align}
where $m_i^2(\phi,T)=m_i(\phi)+\Pi_i$, with $\Pi_i$ being the thermal correction to the mass of the $i$-th boson species. The daisy resummation has to be taken into account because otherwise the bosonic modes with zero Matsubara frequency (which arises due to the finite temperature quantum field theory interpretation) get enhanced in the infrared regime, leading to the breakdown of perturbation theory. It is to be noted that this is the term that generates the term cubic in the field, which in turn gives rise to the temperature-dependent barrier between the true and false minima.
As the temperature of the universe decreases, this temperature-dependent potential acquires a nonzero global minima (true vacuum). The temperature at which the false vacuum (pre-existing local minima) and the true vacuum become degenerate is called the critical temperature ($T_c$). As the temperature continues to lower, the universe shifts to the true vacuum through the nucleation of a true vacuum bubble. The nucleation rate of the true vacuum bubbles can be expressed as,
\begin{align}
\Gamma(T)=\mathcal{A}(T)e^{-S_{3}(T)/T},
\end{align}
where $\mathcal{A}(T)$ is a numerical pre-factor which is $\sim T^4$ and $S_{3}(T)$ is the three-dimensional Euclidean action of the configuration. The Euclidean action is obtained through a Wick rotation ($t\rightarrow-i\tau$), which transforms the oscillatory integral due to the Minkowski action of the form $S = \int d^4x\, \mathcal{L}(\phi,\partial_{\mu}\phi)$ into an exponentially suppressed form as given above where the 4D Euclidean action is given by $S_E = \int d^3x\,d\tau \,\mathcal{L}(\phi,\partial_{\mu}\phi)$. Through the Wick rotation, the tunneling problem becomes analogous to a classical motion in the inverted potential. In finite temperature, the Euclidean time is periodic with period $1/T$ and the thermal tunneling is dominated by time-independent (O(3) symmetric) configurations, which makes the action $S_E = S_3/T$, where the 3D action $S_3$ can be obtained from the expression,
\begin{align}
S_3=\int^{\infty}_{0}dr 4\pi r^2\left[\dfrac{1}{2}\left(\dfrac{d\phi}{dr}\right)^2+V_{\mathrm{tot}}(\phi,T)\right].
\end{align}
The temperature at which the true vacuum bubbles nucleate is called the nucleation temperature $(T_n)$, which can be defined from the idea that at this temperature, the nucleation rate per unit comoving volume is $\mathcal{O}(1)$, i.e.
\begin{align}
\Gamma(T_n)=H^4(T_n),
\end{align}
where $H(T)$ is the Hubble parameter of the universe at temperature $T$ and in a radiation dominated universe it can be approximated as $H(T)\approx 1.66\sqrt{g_*}T^2/M_{\mathrm{Pl}}$ where $g_*$ is the effective degrees of freedom of the universe at temperature $T$ and $M_{\mathrm{Pl}}$ is the Plank mass. Furthermore, the temperature at which $34\%$ of the comoving volume is in true vacuum is referred to as the percolation temperature, $T_p$. This is equivalent to the temperature at which the probability of finding a point in the comoving volume that is still in the false vacuum is 0.71~\cite{Ellis:2018mja}. This probability can be expressed as,
\begin{align}
\mathcal{P}(T)=e^{-\mathcal{I}(T)},
\end{align}
where $\mathcal{I}(T)$ is defined as,
\begin{align}
\mathcal{I}(T)=\dfrac{4\pi}{3}\int^{T_c}_T \dfrac{dT'}{{T'}^4}\dfrac{\Gamma(T')}{H(T')}\left(\int^{T'}_T \dfrac{dT''}{H(T'')}\right)^3.
\end{align}
It is worth mentioning that the strength of a FOPT can be qualitatively understood in a few different ways. If $v_c/T_c>1$ and $v_n/T_n>1$, then the FOPT is said to be a strong FOPT, where $v_c$ and $v_n$ are the finite temperature minima of the field driving the FOPT at $T_c$ and $T_n$, respectively. Furthermore, the ratio between the energy released in the form of latent heat with the radiation energy density at the percolation temperature is also a measure of the strength of the FOPT, which is expressed as,
\begin{align}
\alpha=\dfrac{1}{\rho_{\mathrm{rad}}(T_n)}\left[\Delta V_{\mathrm{tot}}-\dfrac{T}{4}\dfrac{d}{dT}(\Delta V_{\mathrm{tot}})\right]_{T=T_n},
\end{align}
where $\Delta V_{\mathrm{tot}}=V_{\mathrm{tot}}\vert_{\phi=\phi_f}-V_{\mathrm{tot}}\vert_{\phi=\phi_t}$, $\rho_{\mathrm{rad}}(T)=\pi^2 g_* T^4/30$ and $\phi_{f(t)}$ is the value of the field at the false (true) vacuum.
Another property of a FOPT that is of interest to us is its duration, the inverse of which is defined as,
\begin{align}
\frac{\beta}{H} \approx T_n\left[\dfrac{d}{dT}\left(\dfrac{S_3}{T}\right)\right]_{T=T_n},
\end{align}
which suggests that a larger value of $\beta/H$ denotes a faster FOPT. Another quantity of interest is the reheating temperature, i.e., the temperature where the vacuum energy is converted into radiation and this can be expressed as,
\begin{align}
T_{\mathrm{reh}} = T_p(1+\alpha)^{1/4}.
\end{align}
These are the important parameters of a FOPT that enters and takes a pivotal role in the calculation of the GW amplitude and frequency as well as the mass and abundance of the PBH population, which we discuss next.

\section{Primordial Black Holes from First Order Phase Transitions} 
\label{sec:pbhfopt}

First order phase transitions can have several critical phenomenological implications. One such implication is the creation of PBH from FOPTs. It is worth noting that to accommodate PBH creation from FOPT, the FOPT must be both strong and slow. Thankfully, many such models exist where this type of FOPT is possible. Below we discuss the general mechanism of PBH creation briefly before moving onto to the results, which are specific to this mini-review.
\subsection{Model-independent Mechanisms}
\label{sec:genpicpbhfoptpbh}
The creation of PBH requires the formation of overdense regions in the early universe through some mechanism, which may later collapse to form PBH. In this regard, there are two primary mechanisms which can be described in a model-independent manner, i.e. (i) collapse of overdense regions created from delayed vacuum decay (Mechanism-I), and (ii) collapse of overdense regions created due to the curvature perturbations arising from the fluctuation of bubble nucleation time (Mechanism-II). 

In both these mechanisms, the common point is the supercooled nature of the FOPT. In the case of a supercooled FOPT, as the temperature decreases and reaches below the critical temperature ($T_c$), it becomes energetically favorable for the universe to transition to the true vacuum. However, as the temperature decreases further, due to the extremely strong nature of the FOPT, the universe becomes vacuum dominated at the temperature $T_{\mathrm{eq}}~(=\alpha^{1/4}T_n)$. From this temperature to the nucleation temperature, the universe remains in vacuum domination, which leads to a short inflation\footnote{This inflation is different from the inflation which happened in the very early universe.}. After the universe reaches $T_{n}$, the bubbles nucleate in a significant amount, and the vacuum energy is converted into radiation energy, which leads to reheating and the universe becomes radiation dominated, and the temperature increases to the reheating temperature $T_{\mathrm{reh}}$. Following this, the universe again proceeds according to the standard radiation domination.

\subsubsection*{Mechanism-I}
Recently, a few studies have been performed to model the creation of PBH from the collapse of overdense regions created during a FOPT due to delayed vacuum decay~\cite{Liu:2021svg,Kawana:2022olo,Gouttenoire:2023naa,Lewicki:2023ioy,Kanemura:2024pae}. The main idea behind all of these studies are the same, however they differ in some subtleties. The brief outline we provide below is based on Ref.~\cite{Gouttenoire:2023naa}.
Nucleation of true vacuum bubbles, through which a FOPT occurs, is a probabilistic event. Therefore, after nucleation, the percolation of bubbles in one causally connected patch is a random variable. Furthermore, when nucleation of true vacuum bubbles occurs, as these bubbles sweep more and more regions in space, the vacuum energy stored in the false vacuum is released to the expansion of the relativistic bubble walls, as well as thermal and kinetic energy, and scalar waves from bubble collisions. All these different energies then evolve as $a^{-4}$ ($a\equiv$ scale factor of a FLRW universe) and therefore are considered as radiation energy and are termed as $\rho_R$.
Since percolation time is defined as the point at which most of the vacuum energy is converted into radiation energy, the delay in percolation in a patch signifies an overdensity in that patch. This occurs because the patch in question is still dominated by the vacuum energy of the false vacuum and hence remains constant. In contrast, the background is radiation dominated and diluted due to redshift.
The evolution of the radiation energy in a patch can be expressed as,
\begin{align}
    \dot{\rho}_{R}(t;t_{n_i}) + 4H\rho_R(t;t_{n_i}) = -\dot{\rho}_V(t;t_{n_i}),
\end{align}
where $H$ is the Hubble parameter and can be expressed as,
\begin{align}
    H=\sqrt{\dfrac{\rho_V + \rho_R}{3M_{\mathrm{Pl}}^2}},
\end{align}
$M_{\mathrm{Pl}}$ is the Planck mass, and the vacuum energy, which is denoted by $\rho_V$ can be expressed as,
\begin{align}
    \rho_V(t;t_{n_i}) = F(t;t_{n_i})\Delta V,
\end{align}
where $\Delta V$ is the difference in energy between the false and the true vacuum, and $F(t;t_{n_i})$ is the fraction of the volume which is still in false vacuum can in turn be expressed as,
\begin{align}
    F(t;t_{n_i}) = \exp\left[-\int^{t}_{t_{n_i}}dt^{\prime}\Gamma (t^{\prime})a^3(t^{\prime})\frac{4}{3}\pi r^3(t;t^{\prime})\right].
\end{align}
Here $\Gamma$ is the nucleation rate per comoving volume (has been explained in Sec.~\ref{sec:fopt}), $t_{n_i}$ is the time when the first bubble in a patch is nucleated and $r(t;t^{\prime})$ is the comoving radius of a bubble which can in turn be expressed as,
\begin{align}
    r(t;t^{\prime}) = \int^t_{t^{\prime}}\dfrac{d\tilde{t}}{a(\tilde{t})}.
\end{align}
It is to be noted that we denote the total energy density of a delayed patch as $\rho_{\mathrm{tot}}(t;t_{n_i})$ which signifies that in that patch the first bubble nucleates at $t=t_{n_i}$. On the other hand the energy density of the background can be expressed as $\rho_{\mathrm{tot}}^{\mathrm{bkg}}(t) = \rho_{\mathrm{tot}}(t;t_{c})$ where we assume that in the background the bubble nucleation happen at the critical temperature. 
It is also worth-mentioning here that the total energy density of a patch is nothing but the sum of the radiation and vacuum energy density. Now, the overdensity of a delayed patch is quantified as,
\begin{align}
    \delta(t;t_{n_i}) = \dfrac{\rho_{\mathrm{tot}}(t;t_{n_i})-\rho_{\mathrm{tot}}^{\mathrm{bkg}}
    (t)}{\rho_{\mathrm{tot}}^{\mathrm{bkg}}(t)}.
\end{align}

\begin{figure}
    \centering
    \includegraphics[width=0.5\linewidth]{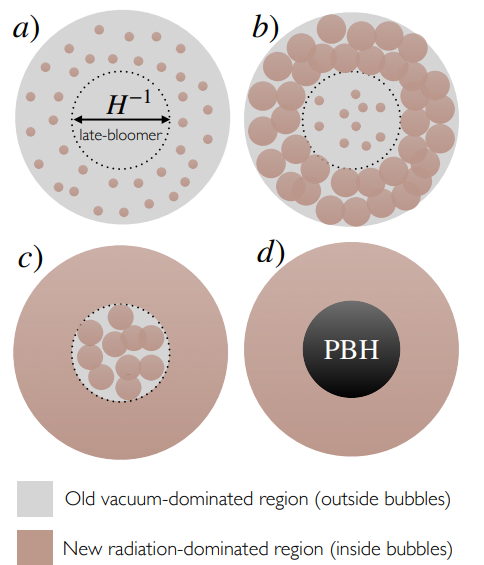}
    \caption{A schematic diagram of the PBH creation mechanism (taken from Ref.~\cite{Gouttenoire:2023naa}).}
    \label{PBHmech1scheme}
\end{figure}

If the value of this overdensity is larger than a certain threshold (mentioned in Sec.~\ref{sec:pbhrecap}), then the region can collapse into a PBH. 
Due to a lack of studies specific to this mechanism, the threshold value has been taken to be $\delta_c = 0.5$. There are a few aspects of this creation mechanism that we express below.
The main mechanism is dependent on the fact that the bubble nucleation process is stochastic. Therefore, if there exists a region in which the nucleation is delayed, then, depending on the delay, the region can eventually collapse into a primordial black hole. This has been expressed chronologically in Fig.~\ref{PBHmech1scheme}.
\begin{enumerate}   
    \item As shown in Fig.~\ref{PBHmech1scheme}, in the top left panel, the delayed patch is initially vacuum dominated and nucleation is yet to occur. In contrast, the surrounding region already has nucleated bubbles.
    \item In the top right panel of the figure, it is prominent that when the bubbles start nucleating in the delayed vacuum ($t=t_{n_i}$), the surrounding region is already almost entirely covered by true vacuum bubbles. This leads to the fact that the radiation energy in the surroundings starts diluting due to redshift, whereas the energy density in the delayed patch dilutes much less.
    \item In the bottom left panel, it is shown that the surroundings have now been completely converted into radiation and hence have been diluted a lot. In contrast, only some part of the energy density of the delayed patch has been converted into radiation, and therefore, the dilution has been less. As a result, there is an overdensity in the radiation energy density in the delayed patch.
    \item The overdensity in the radiation energy density in the delayed patch can lead to a collapse which forms a PBH, as shown in the bottom right panel.
\end{enumerate}
Since the Schwarzschild radius of a Hubble volume is the Hubble horizon ($r_H$), only $0.5$ times overdensity is enough for a collapse. However, for smaller overdense regions, the threshold for collapse is $(r_H/r)^2$ times larger. As a result, for regions much smaller than the Hubble horizon, the threshold is much larger; hence, the probability of collapse of those regions is exponentially suppressed. Hence, only the collapse of Hubble sized regions has been considered.
The probability that an overdense region would lead to an eventual collapse is given by~\cite{Gouttenoire:2023naa},
\begin{align}
    \mathcal{P}_{\mathrm{coll}} = \exp\left[-a\left(\dfrac{\beta}{H}\right)^b(1+\delta_c)^{c\frac{\beta}{H}}\right],
\end{align}
where $\beta/H$ is calculated at the nucleation temperature and $a\approx 1.024$, $b\approx 0.6921$, and $c\approx 0.8831$. It is to be noted that the above expression is valid for $\delta_c \in [0.4,2/3]$ and $\beta/H \in [3,8]$. Furthermore, the abundance of PBHs created through this mechanism can be expressed as~\cite{Gouttenoire:2023naa},
\begin{align}
	f_{\mathrm{PBH}}\approx \left(\dfrac{\mathcal{P}_{\mathrm{coll}}}{2.2\times 10^{-8}}\right)\left(\dfrac{T_{\mathrm{eq}}}{140\mathrm{~MeV}}\right),
\end{align} 
whereas the mass of the PBHs can be expressed as~\cite{Gouttenoire:2023naa},
\begin{align}
	M_{\mathrm{PBH}} \approx \left(\dfrac{20}{g_*(T_{\mathrm{eq}})}\right)^{1/2}\left(\dfrac{140\mathrm{~MeV}}{T_{\mathrm{eq}}}\right)^2M_{\odot},
\end{align}
where $g_*(T)$ is the number of relativistic degrees of freedom in the universe at temperature $T$.
It is evident that in this case the PBHs have a monochromatic mass spectrum since only the collapse of the horizon-sized regions has been considered. Further details of this mechanism can be found in Ref.~\cite{Gouttenoire:2023naa} and references therein.

\begin{figure}[h]
    \centering
    \includegraphics[scale=0.7]{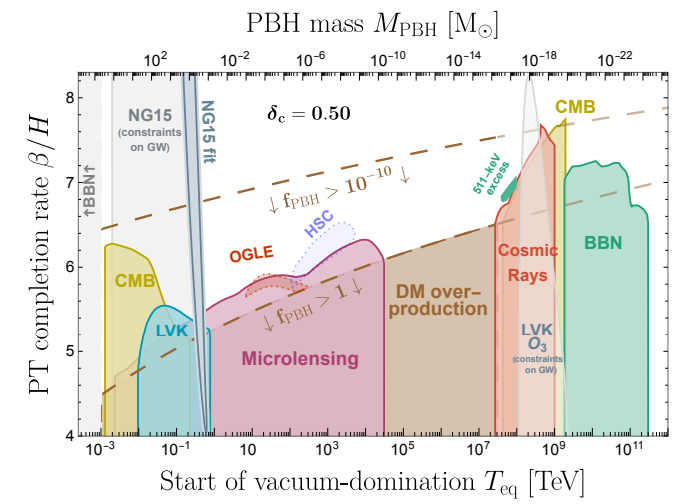}
    \caption{The allowed region of $\beta/H$ and $T_{\mathrm{eq}}$ obtained from this PBH production mechanism. (Taken from Ref.~\cite{Gouttenoire:2023naa}.)}
    \label{fig:plotgv}
\end{figure}
In Fig.~\ref{fig:plotgv} the authors use the existing bounds on the PBH mass-abundance plane (discussed in Sec.~\ref{sec:pbhrecap}) and convert them into the bounds on the $T_{\mathrm{eq}}-\beta/H$ plane. Furthermore, they have also used the constraints on the stochastic gravitational wave background, which is a direct consequence of a strong FOPT, to put bounds on the $T_{\mathrm{eq}}-\beta/H$ plane. Hence, if this mechanism is taken into account, then the model parameters are to be chosen such that the resulting $T_{\mathrm{eq}}$ and $\beta/H$ have to be in the allowed region of the above figure.
\subsubsection*{Mechanism-II}
This mechanism has the same basic principle but the dynamics and the process leading to the creation of PBH is different than the one discussed before.
In this case also, the primary focus lies on the delay of the nucleation of true vacuum bubble in different patches. However, unlike the previous case, here the difference in the time of bubble nucleation in different patches is directly translated into the curvature perturbation in the superhorizon regime. 
Below we briefly outline the entire mechanism, where the creation of the curvature perturbation is based on Ref.~\cite{Elor:2023xbz} and the creation of PBH from these curvature perturbations is based on Ref.~\cite{Lewicki:2024ghw}.
As mentioned above, the bubble nucleation is a stochastic process. As a result, the time of bubble nucleation at different patches of the universe may differ from each other. Hence, if there are two points in the universe that transition from false vacuum to true vacuum at different times, due to the expansion of the universe, the point that transitions first dilutes more than the other and a difference in the radiation energy density is created.  
The difference in the time of transition is defined as,
\begin{align}
\delta t_c(\vec{x}) = t_c(\vec{x})-\overline{t}_c,
\end{align}
where $t_c(\vec{x})$ is the time of transition of the point $\vec{x}$ and $\overline{t}_c$ is the average time of transition. In order to statistically estimate the effect of this delay, it is imperative to use the two-point correlation function $\langle \delta t_c(\vec{x})\delta t_c(\vec{y})\rangle$. 
Furthermore, since we eventually use the power spectrum of the curvature perturbation, the Fourier transformation of this two-point correlation function is crucial and is defined as~\cite{Elor:2023xbz},
\begin{align}
\mathcal{P}_{\delta t}(k) = \dfrac{k^3}{2\pi^2}\left(\dfrac{H}{\beta}\right)\int d^3 r e^{i\vec{k}\cdot \vec{r}}\beta^2\langle \delta t_c(\vec{x})\delta t_c(\vec{y})\rangle, 
\end{align}
where $\vec{r}=\vec{x}-\vec{y}$, i.e., the distance between two points in space. At this point, it becomes important to divide the problem into two different cases based on the scales, i.e., (i) scales which are smaller or equal to the bubble size and (ii) scales which are much larger than the bubble size. 
It is to be noted here that the bubble size is usually defined as $v_b/\beta$ where $v_b$ is the velocity of the bubble wall. For the first case, it is difficult to calculate the final expression of the time delay in a model independent manner as the microphysics becomes important. On the other hand, for the second case, using the central limit theorem, one can express $\mathcal{P}_{\delta t}$ as~\cite{Elor:2023xbz},
\begin{align}
	\mathcal{P}_{\delta t} \approx 70.4 v_b^3\left(\beta/H\right)^{-5}\left(\dfrac{k}{\mathcal{H}}\right)^3,
\end{align}
where $\mathcal{H} (= aH)$ is the comoving Hubble parameter. The most important feature which can be immediately seen form the above expression is the dependence of $\beta/H$ and $k$.  The power spectrum of the curvature perturbation created from this time delay can be expressed as~\footnote{In Ref.~\cite{Elor:2023xbz}, the authors concerned themselves with phase transitions in the dark sector. However, here since we discuss general phase transition, the expression of the curvature power spectrum has been modified.}~\cite{Elor:2023xbz},
\begin{align}
	\mathcal{P}_{\zeta}(k) = \dfrac{1}{4}\left(\dfrac{\alpha}{1+\alpha}\right)^2\mathcal{P}_{\delta t}(k).
    \label{eq:curvpertrj}
\end{align}
These curvature perturbations, in turn, lead to a non-Gaussian density distribution. More precisely, the distribution of the overdensity can be expressed as~\cite{Lewicki:2024ghw},
\begin{align}
	P_{k}(\delta) \propto \exp\left[\dfrac{\epsilon}{2}(\delta-\mu) - \dfrac{2}{\epsilon^2\sigma^2}\left(1-e^{\frac{\epsilon}{2}(\delta-\mu)}\right)^2\right],
\end{align}
where $\epsilon,~\mu>0$ which signifies an exponential behaviour for $\delta<0$ and a much rapid decrement for $\delta>0$. This non-Gaussianity and non-monochromaticity leads to an extended mass function of the eventual PBHs. Therefore, unlike the previous case, there is no single mass of PBH population, rather its a distribution. However, the mass distribution of a PBH population is closely linked to the horizon mass at the time of collapse.
The horizon mass during the collapse of the overdense region can be expressed as,
\begin{align}
M_H = 10^{32}\left(\dfrac{100}{g_*}\right)^{1/2}\left(\dfrac{T_{\mathrm{reh}}}{\mathrm{GeV}}\right)^{-2} \mathrm{g}.
\end{align}
The abundance of the PBH created through this mechanism is given by~\cite{Lewicki:2024ghw},
\begin{align}
f_{\mathrm{PBH}} \approx 5.5\times 10^6 \exp\left(-0.064e^{0.806\beta/H}\right)\left(\dfrac{g_*}{g_{*s}}\right)\left(\dfrac{T_{\mathrm{reh}}}{\mathrm{GeV}}\right).
\end{align}
Furthermore, the mass function for these PBHs can be expressed as~\cite{Lewicki:2024ghw},
\begin{align}
\psi(M) \propto (M/M_H)^{1+1/\gamma}\exp\left(-c_1(M/M_H)^{c_2}\right),
\end{align}
where $\gamma$ is the coefficient of the critical scaling law of the PBHs which takes the form $M = \kappa M_k(\delta-\delta_c)^{\gamma}$ where $\kappa$ is the numerical prefactor, $M_k$ is the horizon mass when the perturbation enters the horizon, $\delta$ is the overdensity of the region and $\delta_c$ is the overdensity threshold beyond which a region collapses as the radiation pressure can not balance the overdensity. In this case, fixed values of $\gamma = 0.36$, $\kappa = 4$, and $\delta_c = 0.5$ have been considered. It is also to be noted that the values $c_1$ and $c_2$ have weak dependence on $\beta/H$ and they take approximate values of $1.2$ and $2.7$ respectively for $\beta/H = 8$~\cite{Lewicki:2024ghw}. Therefore, it is trivial to see that in case of a FOPT of this type, the maximum PBHs of a certain population will have a certain mass $M_{\mathrm{PBH,~peak}} \approx 0.928M_H$, which we term as `peak mass'.
\begin{figure}[h]
    \centering
    \includegraphics[scale=0.7]{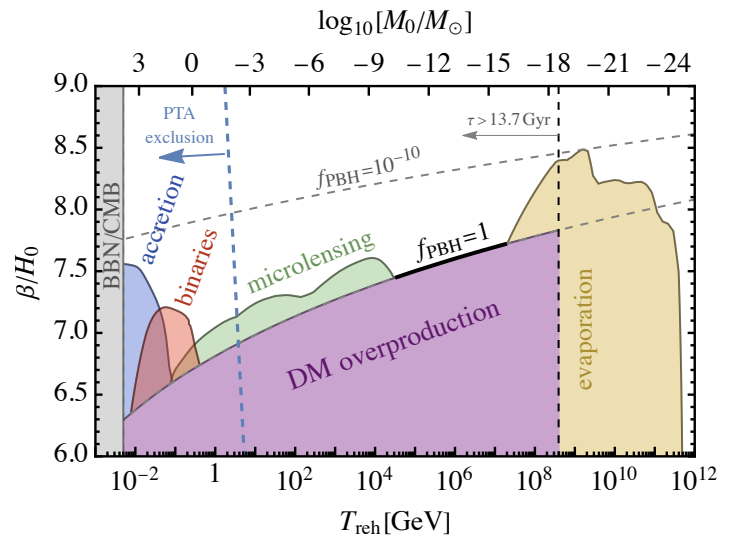}
    \caption{The allowed region of $\beta/H$ and $T_{\mathrm{reh}}$ obtained from this PBH production mechanism. (Taken from Ref.~\cite{Lewicki:2024ghw}).}
    \label{fig:plotlt}
\end{figure}
Similar to Fig.~\ref{fig:plotgv}, here also the authors convert the PBH abundance bounds and the gravitational wave constraints to bounds on the $T_{\mathrm{reh}}-\beta/H_0$ plane, which we show in Fig.~\ref{fig:plotlt}. It is to be noted here that in this mini-review, we term the PT completion rate (inverse duration of the PT) as $\beta/H$ whereas in Ref.~\cite{Lewicki:2024ghw}, the authors term this quantity as $\beta/H_0$ and both these quantities carry the same significance.
\subsection{Model-dependent Mechanisms}
Apart from the model-independent mechanisms mentioned above, there are other possibilities of formation of PBH from FOPT, but they depend on specific models. Some of these mechanisms are described briefly below.
\subsubsection*{Fermi Ball}
Fermi balls are hypothetical bound astrophysical objects. The gravitational inward pressure in a Fermi ball is countered by the degeneracy pressure of the fermions which create Fermi balls. These Fermi balls have interesting phenomenological implications, i.e., they can be extended dark matter objects or they can be the intermediate steps toward the creation of black holes~\cite{Kawana:2021tde}. We will focus on the second one in this mini-review.
During a FOPT, Fermi balls (FB) may be generated from the entrapment of fermions in false vacuum regions. There are three main conditions which are necessary for Fermi balls to be created during a FOPT,
\begin{enumerate}
    \item The mass of the fermions in the false vacuum region must be significantly less than the mass of the fermions in the true vacuum. As a result, during the FOPT when the true vacuum bubbles are nucleated, a large fraction of these fermions, which initially stay in false vacuum, do not have enough momentum to cross the bubble wall to proceed to the true vacuum. This paves the way for the entrapment of the fermions in the false vacuum fraction.
    \item There must be an asymmetry between the fermions and anti-fermions such that the trapped fermions can not annihilate completely into other species of particles. This results in the excess fermions creating the Fermi balls.
    \item There should be a global $U(1)$ charge carried by the fermions. This makes the Fermi balls acquire a finite charge under the $U(1)$ symmetry, making it stable.
\end{enumerate}
In a model-dependent scenario there are several caveats that one needs to take care in order to get the entrapment of the fermions followed by the collapse to form PBHs. Below we briefly explain the requirements for the same.
\begin{itemize}
    \item \textbf{Particle content:} The basic mechanism requires a few additions to the standard model, i.e., a fermion species ($\chi$) with asymmetry, a scalar field ($\phi$) which acquires a vacuum expectation value (VEV) as the universe cools down, leading to a FOPT.
    \item \textbf{Adequate coupling:}  Furthermore, there must exist an adequate Yukawa coupling $y_{\chi\phi}$ between $\phi$ and $\chi$, and in the false vacuum, the mass of $\chi$ is very low. As the universe cools down and $\phi$ acquires a VEV $w$, in the true vacuum $\chi$ acquires an additional mass $M_{\chi}=y_{\chi\phi}w$, which is much larger than the bare mass of $\chi$.
    \item \textbf{Transition temperature and wall velocity:} If the transition temperature is $T_{*}$, then under the condition of $M_{\chi}\gg T_*$ these $\chi$ particles get trapped in the false vacuum as they do not posses energy high enough to pass through the true vacuum bubble wall. In this regard, it is to be noted that the bubble wall velocity $v_b$ plays a crucial role, as for higher $v_b$ it is easier for $\chi$ particles to pass through the bubble wall. In Fig.~\ref{fig:fb_trap_frac}, we show the dependence of the fraction of the fermions that get trapped ($F_{\chi}^{\mathrm{trap}}$) on the ratio $M_{\chi}/T_*$ for different wall velocities $v_b$.
    \begin{figure}[h]
        \centering
        \includegraphics[width=0.7\linewidth]{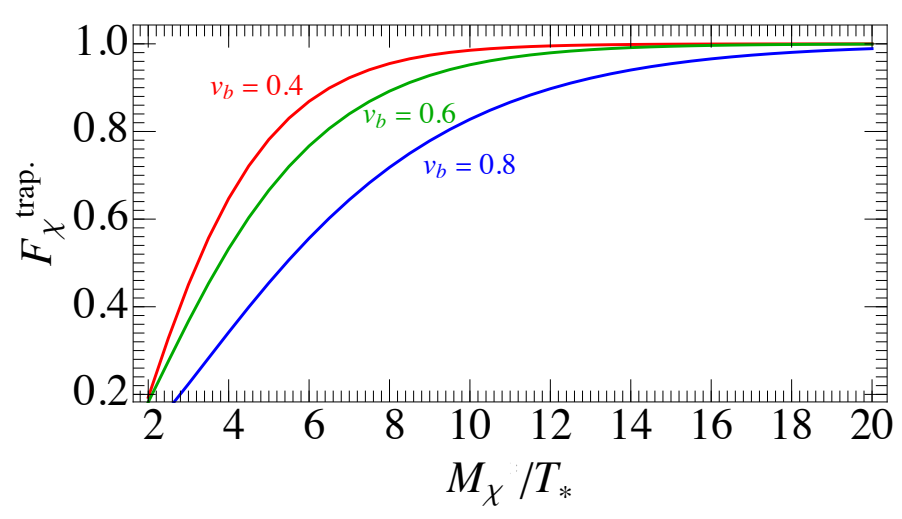}
        \caption{The dependence of fraction of fermions that get trapped ($F_{\chi}^{\mathrm{trap}}$) with respect to the ratio $M_{\chi}/T_*$ for different bubble wall velocities $v_b$. This figure is taken from Ref.~\cite{Hong:2020est}.}
        \label{fig:fb_trap_frac}
    \end{figure}
    It can be seen from Fig.~\ref{fig:fb_trap_frac} that trapping fraction is negligible for $M_{\chi}/T_* < 2$, on the other hand for $M_{\chi}/T_* > 20$ trapping fraction reaches unity. For the intermediate values of $M_{\chi}/T_*$ the trapping fraction reduces drastically with increasing values of bubble wall velocity. 
\end{itemize}
  It is also to be noted here that the strength $\alpha$ and the inverse duration $\beta/H$ also plays a crucial role in determining the fraction of the universe in which the $\chi$ particles are trapped. These regions, where the $\chi$ particles are trapped, are called Fermi balls. Now as the $\chi$ particles have some intrinsic asymmetry, all the $\bar{\chi}$ get annihilated, leaving only $\chi$ particles behind.

\begin{figure}
    \centering
    \includegraphics[width=0.95\linewidth]{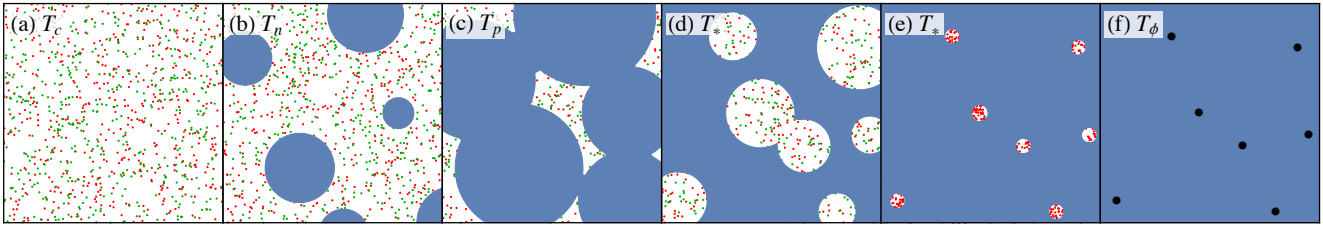}
    \caption{The schematic diagram of the PBH creation mechanism for the collapse of Fermi balls. The white (blue) region is the false (true) vacuum region. (a) The $\chi$ particles ($\overline{\chi}$ antiparticles) are shown in red (green) dots in the false vacuum. (b) The true vacuum bubbles start nucleating, but due to the high mass difference, a large fraction of the fermions are unable to enter the true vacuum. (c) Most of the fermions are trapped in the false vacuum region as more and more fraction of the universe transitions to true vacuum. (d) Separate structures of these trapped fermions are formed. (e) Due to the asymmetry between the $\chi$ and $\overline{\chi}$, the $\chi\overline{\chi}$ annihilation takes place and only the $\chi$ particles get left behind. This is the creation of Fermi balls. (f) Finally, after the Fermi balls cool down, the Yukawa interaction overpowers and leads to the collapse of the Fermi balls in PBHs. The figure is taken from Ref.~\cite{Kawana:2021tde}.}
    \label{fig:FB_cartoon}
\end{figure}

Once the Fermi balls form, they cool down as the $\chi$ radiate lighter particles. At this stage, the Fermi balls reach a temperature where the Yukawa force between the $\chi$s dominates and the Fermi balls collapse into a PBH. The entire mechanism has been shown in Fig.~\ref{fig:FB_cartoon} through a schematic cartoon.

The PBH mass and abundance can be expressed as~\cite{Kawana:2021tde},
\begin{align}
M_{\mathrm{PBH}} &= 1.4\times 10^{21} \left(\dfrac{\eta_{\chi}}{10^{-3}}\right)\left(\dfrac{100}{g_*}\right)^{1/4}\left(\dfrac{100\mathrm{~GeV}}{T_*}\right)^2\left(\dfrac{100}{\beta/H}\right)^3\alpha^{1/4} v_b^3~\mathrm{~g},\nonumber \\
f_{\mathrm{PBH}} &= 1.3\times 10^3 \left(\dfrac{g_*}{100}\right)^{1/2}\left(\dfrac{T_*}{100\mathrm{~GeV}}\right)^3\left(\dfrac{\beta/H}{100}\right)^3v_b^{-3}\left(\dfrac{M_{\mathrm{PBH}}}{10^{15}\mathrm{~g}}\right),
\end{align}
where $g_*$ is the effective degrees of freedom at $T_*$ and $\eta_{\chi}$ is the asymmetry of $\chi$ and it can be expressed as,
\begin{align}
\eta_{\chi} &= \dfrac{n_{\chi}-n_{\bar{\chi}}}{s},
\end{align}
where $n_{\chi}$ ($n_{\bar{\chi}}$) are the number densities of $\chi$ ($\bar{\chi}$) and $s=2\pi^2g_*T^3/45$ is the entropy density of the universe. It is worth mentioning here that this is a very brief description of Fermi ball PBH, and the reader is encouraged to go through Refs.~\cite{Hong:2020est,Kawana:2021tde} for further details.

\subsubsection*{Q-Ball}
Q-balls are non-topological solitons, which are a localized field configuration of complex scalar fields. These are usually stable due to the conserved charge under a global symmetry.

Similar to the Fermi balls, Q-balls can also be created from the entrapment of particles in false vacuum regions of the universe during a FOPT. However, in this case the trapped particles are long-lived bosons and hence some of the dynamics differ from the previous case. The main steps of the fomation of PBH from the collapse of Q-balls can be expressed as follows~\cite{Dent:2025lwe}.
\begin{itemize}
    \item \textbf{Mass and asymmetry:} In accordance with the previous case, here also the mass of the boson has to be much larger than in the true vacuum than that of in false vacuum to facilitate the entrapment. Furthermore, the asymmetry between the bosonic particle and antiparticle is also required for the same reason as that of in the case of Fermi balls.
    \item \textbf{Formation of Q-balls:} Once these bosonic particles are trapped in the false vacuum regions, the regions continue to shrink until a difference in pressure exists between the true and false vacuum regions. Once that pressure difference vanishes, the regions stabilize. Since in this case the particles are bosons, once the temperature drops below the Bose-Einstein condensation temperature, most of the particles crowd the ground state, creating a Q-ball. This is the main difference between the Fermi balls and Q-balls. Due to this reason, the Q-balls have much weaker pressure to balance the gravitational collapse and that leads to the Q-balls being much smaller than Fermi balls before the collapse.
    \item \textbf{Collapse of Q-balls -- Yukawa-type attraction:} These Q-balls can then collapse to form a primordial black hole due to some Yukawa-type attractive force between the bosonic particles. Initially, when the temperature is high enough, the interaction is short-range due to the thermal corrections in the mass of the mediator particle. But as the temperature goes down, the effective mass of the mediating particle reduces, and the Yukawa interaction becomes important. If the strength of this interaction is higher than a certain threshold, the Q-balls collapse to form a PBH. As mentioned before, Q-balls can have a much smaller size in comparison to the Fermi balls before the collapse, and hence the mass of the PBH that originates from the collapse of Q-balls can be much smaller than that of the ones originating from Fermi balls.
\end{itemize}

The properties of these black holes depend on the specifics of the model. The initial mass of PBH can be obtained from the free energy of a Q-ball at the time of formation, which can be expressed as~\cite{Dent:2025lwe},
\begin{align}
    M_{\mathrm{PBH}}\approx F_{\mathrm{collapse}} \approx 1.65\pi\tilde{U}_0^{1/4}(T)Q^{3/4},
\end{align}
with $\tilde{U}_0(T)=U_0(T)-(\pi^2/90)T^4$, where $U_0(T)$ is the difference between the true and false vacuum energy density at temperature $T$, and $Q$ denotes the number of particles in a false vacuum region of radius $R_f$ at the time of formation and can be expressed as~\cite{Dent:2025lwe},
\begin{align}
    Q=\dfrac{\eta_{b}s(T_p)}{F(T_p)}\dfrac{4\pi}{3}R_f^3,
\end{align}
where $F(T_p)$ is the fraction of the universe which is in false vacuum at the time of percolation, which by definition is $F(T_p)=1/e$, and the value of $R_f$ can be assumed to be equal to $R_{f_p}=(3v_b/4\pi\Gamma_p)^{1/4}$ where monochromatic bubble size is assumed. 
In case the bubble size is not considered to be monochromatic, the bubbles follow a distribution where the number remains constant for $R_f<R_{f_p}$ but falls steeply for $R_f>R_{f_p}$. Here, $\Gamma_p$ is the rate of nucleation of true vacuum bubbles at the percolation temperature.
The abundance of these PBHs can be expressed as~\cite{Dent:2025lwe},
\begin{align}
    f_{\mathrm{PBH}} = 1.7\times 10^{6}\dfrac{T_p}{\mathrm{MeV}}\dfrac{M_{\mathrm{PBH}}}{\rho_{\mathrm{tot}}(t_i)}\dfrac{F(T_p)}{(4\pi/3)R_f^3},
\end{align}
where $\rho_{\mathrm{tot}}(t_i)$ is the total density of the universe at the time of formation $t_i$.

\subsubsection*{Cosmic Strings}
Cosmic strings are one dimensional topological defects which may have been created during the breaking of a continuous symmetry in the early universe~\cite{Kibble:1976sj}. Therefore, these strings might be a direct consequence of first order phase transitions in various BSM models where a local U(1) symmetry breaks due to a scalar acquiring a VEV. After the local symmetry is broken and most of the universe shifts into the true vacuum, these strings arise as one-dimensional solutions which remain in the false vacuum. As a result, the cosmic strings may contain a large amount of energy (due to the difference in energy between true and false vacua) which may lead to various phenomenological implications.
Usually, these strings are characterized by their tension $\mu$, which is also expressed in terms of energy per unit length of the string. Although the specific nature of the scalar potential is required to determine the exact value of the string tension, an estimate can be provided using the scale of the symmetry breaking, i.e., $\mu\sim T^2$, where $T$ is the scale of the symmetry breaking. Furthermore, for simplification, the string tension is often expressed as the dimensionless quantity $G\mu$, where $G$ is Newton's constant\footnote{The string tension is expressed as $G\mu$ in natural units where $\hbar=c=k_B =1$ whereas in SI units, this is expressed as $G\mu/c^2$.}.
These strings are virtually of infinite length when formed, but as the universe expands, more and more regions of these strings enter the Hubble regions and make way for these strings to intersect and intercommute. As a result, the strings can then form loops. These loops can radiate energy in the form of gravitational waves while oscillating. A fraction of these cosmic string loops may collapse to form PBH if they contract to a size smaller than their Schwarzschild radius. This contraction takes place as the loops succumb under their own tension~\cite{Hawking:1987bn,Polnarev:1988dh}.
In order to calculate the properties of PBHs formed from the collapse of cosmic string loops, one needs to calculate the rate of formation of cosmic string loops, which can be expressed as~\cite{James-Turner:2019ssu},
\begin{align}
   \dfrac{dn_l}{dt} = \dfrac{A V(t) \delta}{8\alpha_l c^3 t^4}\left(1-\langle v^2\rangle \right),
\end{align}
where $n_l$ is the number of cosmic string loops, $A$ denotes a numerical prefactor related to the density of the infinite strings, $\delta$ is the fraction of the energy which lies in the large loops\footnote{The large loops are the ones which may eventually collapse to form PBH whereas the smaller loops do not contribute to the final PBH population.}, $V(t)$ is the Hubble volume at cosmic time $t$, $\langle v^2 \rangle$ is the mean squared velocity of the infinite strings, $\alpha_l$ is the ratio between the loop length and the horizon length. Recent numerical simulations indicate values of the constants as $A=44$, $\alpha_l\sim 0.1$, and $\delta\sim 0.1$~\cite{Blanco-Pillado:2011egf}. Furthermore, it has been shown that in radiation dominated universe, $\langle v^2 \rangle \sim 0.4$~\cite{Blanco-Pillado:2011egf}.
The rate of PBH formation can be expressed as,
\begin{align}
\dfrac{dn_{\mathrm{PBH}}}{dt} = f\dfrac{dn_l}{dt},
\end{align}
where $n_{\mathrm{PBH}}$ denotes the number of PBHs and $f$ denotes the fraction of the large loops that collapse to form PBH. It is to be noted here that the value of $f$ is yet undetermined for realistic cases of string tension but there exists some constraints on $f$. Finally, on can express the PBH abundance as,
\begin{align}
    f_{\mathrm{PBH}} = \dfrac{1}{\Omega_{\mathrm{CDM}}~\rho_{c,0}}\int^{t_0}_{t_*}\dfrac{dn_{\mathrm{PBH}}}{dt^{\prime}}M(t^{\prime})dt^{\prime},
\end{align}
where $\rho_{c,0}$ is the current critical density of the universe, $M(t^{\prime}) = \alpha_l\mu ct^{\prime}$ denotes the mass of a PBH which formed at time $t^{\prime}$, and $t_*$ is the time of formation of lightest PBH which has not yet evaporated completely, i.e., PBH of mass $M_*\approx 5\times 10^{14}\mathrm{~g}$.
Another interesting properties of the PBHs which are formed through the collapse of cosmic string loops is that their mass function is not monochromatic, rather it follows a scaling relation of the form $dn_{\mathrm{PBH}}/dM \propto M^{-5/2}$~\cite{James-Turner:2019ssu}, which suggests that the lightest non-evaporated PBHs will have the highest population.
Although the collapse of cosmic string loops is a viable creation mechanism for PBHs, mainly because of the abundance of models which predict the generation of cosmic strings in sufficient abundance, the uncertainties pertaining to some of the quantities (e.g. $f$) makes this mechanism less studied using specific models. Hence in the next section we do not provide any model specific examples for this mechanism.

\section{Examples}
\label{sec:example}
In this section, we discuss a few works as examples of how these various mechanisms have been used in literature.

\vspace{0.3cm}
\noindent
\textbf{Model-independent mechanism - I:} In Ref.~\cite{Gouttenoire:2023pxh}, the authors have considered a conformal Higgs paradigm, i.e., they have extended the SM with  a complex scalar $\Phi$ which is not charged under the SM gauge group but rather has a charge under a BSM gauge group $U(1)_D$. In this work, the potential has been taken to be scale-invariant and therefore it takes the form,
\begin{align}
    V_{\mathrm{tree}} = \lambda_h |H|^4 + \lambda_{\Phi}|\Phi|^4 + \lambda_{h\Phi}|H|^2|\Phi|^2,
\end{align}
where $H$ is the SM Higgs. It is to be noted that due to the absence of the mass term for the field $\Phi$, the zero-temperature VEV is realized through the one-loop quantum corrections. 
Once the field $\Phi$ acquires a VEV $v_\Phi$, the term $\lambda_{h\Phi}|H|^2|\Phi|^2$ generates the SM Higgs mass term eventually leading to the tree-level minima of the Higgs potential. The authors have used the value of Higgs VEV $v_{\mathrm{EW}} = 246$ GeV and the Higgs mass $m_h = 125$ GeV to find a relation between the coupling $\lambda_{h\Phi}$ and $v_\Phi$. In this kind of a model, where the zero-temperature VEV is generated through the one-loop quantum corrections, the FOPT is usually very strong and sufficiently slow, which is suitable for this mechanism. The authors find that if the Higgs mixing is very small ($\lambda_{h\Phi}\ll 1$) then the $v_\Phi \in 4\times [10^2, 10^5]\mathrm{~TeV}$ can generate PBHs which can account for the entire dark matter of the universe.
\vspace{0.3cm}
\noindent
\textbf{Model-independent mechanism - II:} In Ref.~\cite{Banerjee:2024fam}, this mechanism has been used for a $U(1)_{B-L}$ extended inert doublet model (IDM). In this model, there are three scalars; an SM singlet $\chi$ which acquires a VEV and spontaneously breaks the $U(1)_{B-L}$ symmetry. The other two scalars $\Phi_{1,2}$ are doublet under $SU(2)_L$ symmetry, and the $\Phi_1$ eventually breaks the electroweak symmetry. It is to be noted that we impose an additional $\mathbb{Z}_2$ symmetry under which the $\Phi_1$ is even, and the $\Phi_2$ is odd, and hence the latter acts as an inert doublet.  
Furthermore this model also has three heavy right handed neutrinos that can generate the active neutrino mass through the seesaw mechanism. This model also has scale-invariance and hence the tree-level potential takes the form,
\begin{align}
V(\Phi_1, \Phi_2, \chi) = &\lambda_1 \vert \Phi_1\vert^4 + \lambda_2 \vert \Phi_2 \vert^4 +  \lambda_3 \vert\Phi_1\vert^2 \vert\Phi_2\vert^2 +  \lambda_4 \vert \chi \vert^4 +  \lambda_5 \vert \Phi_1^\dagger \Phi_2 \vert^2  \nonumber\\
&+ \lambda_6 \left[(\Phi_1^\dagger \Phi_2)^2 + \text{h.c.}\right]  +\  \lambda_7 \vert \chi \vert^2  \vert \Phi_1\vert^2 + \lambda_8  \vert \chi \vert^2  \vert \Phi_2\vert^2 ,
\label{Vtree}
\end{align}
where $( \lambda_1,\ldots, \lambda_8 )$ are dimensionless coupling constants which determine the strength of the interactions. Due to the scale-invariance the field $\chi$ cannot acquire a VEV solely from the tree-level potential and the VEV $v_{\chi}$ is generated from the one-loop quantum corrections. This VEV then generates the mass terms for the fields $\Phi_{1,2}$ and eventually leads to the electroweak symmetry. It has been shown for a few benchmark cases that PBHs can be obtained in this model which can account for the entire dark matter of the universe for $v_\chi \in [10^7,10^8]\mathrm{~GeV}$ where the Yukawa coupling between the right handed neutrinos and the field $\chi$ is $\mathcal{O}(0.1-0.2)$, and the  $U(1)_{B-L}$ gauge coupling is $\mathcal{O}(0.1-0.2)$.
\vspace{0.3cm}
\noindent
\textbf{Fermi Ball:} The Fermi ball PBH scenario has been adopted in Ref.~\cite{Borah:2025wzl}, where an additional Affleck-Dine scalar $\eta$ has been considered, which is has a global $U(1)_B$ charge 2, but remains a singlet under the SM gauge. Additionally, an extra Dirac fermion $\chi$ has been considered, which carries a global charge of 1. 
There exists a Yukawa interaction term between the fermions and the scalar of the form $Y_D\overline{\chi^c}\chi\eta^{\dagger}$, which provides the required attractive force for the Fermi balls to collapse into PBH. Furthermore, the asymmetry necessary between the fermions and the anti-fermions translates from the asymmetry in $\eta$ via the decay channel $\eta\rightarrow \chi\chi$, where the asymmetry in $\eta$ is generated from a term $\epsilon m_\eta^2\eta^2$ which explicitly breaks the $U(1)_B$. In order to maintain the significant difference in the mass of $\chi$ in true and false vacuum, the authors introduce a singlet scalar $\Phi_1$, which eventually drives the FOPT. Furthermore, the authors consider a heavy scalar $\zeta$ to transfer the asymmetry in $\chi$ to some SM fermion $\psi$ and another singlet scalar $\Phi_2$ to assist in the FOPT.
Within this framework, the authors show a few benchmark cases where FOPT gives rise to Fermi ball PBH with $f_{\mathrm{PBH}}=1$, which shows that this kind of scenario can lead to the PBHs that can account for all the dark matter in the universe while also leading to other interesting implications, such as testable gravitational waves and explanation of baryon asymmetry.
\vspace{0.3cm}
\noindent
\textbf{Q-Ball:} PBH production from the collapse of Q-ball has been investigated in Ref.~\cite{Dent:2025lwe}, where the authors consider a scenario where the FOPT is driven by a scalar field $\phi$ and it couples with a boson $X$ which acquires a large mass term in the true vacuum due to the VEV of the $\phi$ field. The effective thermal potential in this case is given by,
\begin{align}
    U(\phi,T) = \Lambda^4\left[\left(-\dfrac{1}{2}+c\left(\frac{T}{c}\right)^2\right)\left(\frac{\phi}{v}\right)^2+b\frac{T}{v}\left(\frac{\phi}{v}\right)^3+\frac{1}{4}\left(\frac{\phi}{v}\right)^4\right]+\Lambda_0^4,
\end{align}
where the value of $\Lambda_0^4$ is set such that the vacuum energy density of the true vacuum is the cosmological constant. The quantities $b,\,c$ are dimensionless constants such that, $c>0,~b<0,~c/b^2>1$. The critical temperature corresponding to the potential $T_c = v/\sqrt{2(c-b^2)}$. For $T>T_c$, the high temperature global minima is at $\phi=0$ which becomes the false minima at low temperature $T<T_c$ and the true vacuum becomes,
\begin{align}
    \phi_{\mathrm{true}} = \frac{1}{2}\left(-3bT+v\sqrt{9b^2(T/v)^2+4(1-2c(T/v)^2)}\right).
\end{align}
The difference in the energy density between the true and false vacuum in this set-up can be expressed as,
\begin{align}
    U_0(T) = \Lambda_0^4 - U(\phi_{\mathrm{true}},T).
\end{align}
In this scenario, the dark sector ($\phi,~X$) are decoupled from the SM, and hence the temperature of the SM plasma and the dark sector plasma can differ. This difference is expressed as a ratio, and at the percolation, this ratio can be expressed as $\zeta_* = T_{\mathrm{SM}*}/T_*$. 

In this scenario, the authors show that to evade constraints from the Hawking evaporation of the PBH while making sure that the PBH, and GW from the FOPT will be observed in the upcoming experiments, the percolation temperature has the bound $100\mathrm{~keV}\lesssim T_*\lesssim 1\mathrm{~MeV}$. The lower bound comes from the fact that FOPTs with lower percolation temperature makes the abundance of the PBH too small for future detection. The upper bound comes from the fact that PBHs formed from FOPT in the higher energy will have very small mass, and hence they are constrained by the Hawking evaporation bounds.

\section{Conclusion}
\label{sec:conc}
In this mini-review, we discuss the creation of PBHs from various model-(in)dependent mechanisms, during a FOPT. 
We start with a brief description of the general PBH creation mechanism and the various constraints on it. We then describe the FOPT mechanism in detail and discuss its important properties, which are relevant to the PBH creation scenario. We discuss the PBH creation mechanisms, where in the model-independent scenario, we focus on two processes, i.e., (i) collapse of overdense regions created from the delayed vacuum decay during a slow and strong FOPT, (ii) collapse of overdense regions created from the curvature perturbations originating from the delay in nucleation time during a strong and slow FOPT. Next, we discuss a few model-dependent mechanisms where PBHs are created indirectly from (i) Fermi balls, (ii) Q-balls, and (iii) cosmic strings, which are created during a FOPT. 
Finally, we provide a few model-dependent examples for each of these mechanisms apart from the one from cosmic strings. This is due to the fact that in the mechanism related to cosmic strings, there are various uncertainties related to concerned quantities, such as the collapse fraction, which renders it difficult to use in realistic model-dependent scenarios. With improved simulations in the future, when the uncertainties related to this mechanism are mitigated to some extent, they can be utilized for specific model building.
This mini-review serves as a comprehensive collection of the brief description of PBH creation mechanisms during a FOPT. It is worth-noting that there are various articles related to each of the creation mechanisms discussed in this mini-review. Although we have cited them, we have not referenced each of them in the discussions. The interested reader is encouraged to go through the references to understand the various aspects of each of these mechanisms.

\acknowledgments{
UKD acknowledges support from the Anusandhan National Research Foundation (ANRF), Government of India under Grant Reference No.~CRG/2023/003769. The work of SK is partially supported by Science, Technology \& Innovation Funding Authority (STDF) under grant number 48173.}
\bibliographystyle{JHEP}
\bibliography{ref.bib}

\end{document}